\documentclass[prb,aps,nofootinbib,preprintnumbers]{revtex4}
\usepackage{amsmath,amssymb,epsf}
\usepackage{graphicx}

\begin{document}
\title{CurvedLand: An Applet for Illustrating Curved Geometry without Embedding}
\author{Gary Felder and Stephanie Erickson}
\affiliation{Department of Physics, Clark Science Center, Smith College Northampton, MA 01063, USA}

\begin{abstract}
We have written a Java applet to illustrate the meaning of curved geometry. The applet provides a mapping interface similar to MapQuest or Google Maps; features include the ability to navigate through a space and place permanent point objects and/or shapes at arbitrary positions. The underlying two-dimensional space has a constant, positive curvature, which causes the apparent paths and shapes of the objects in the map to appear distorted in ways that change as you view them from different relative angles and distances.
\end{abstract}

\maketitle

\section{Introduction}

\begin{figure} [h]
\begin{center}
\includegraphics[width = 90mm]{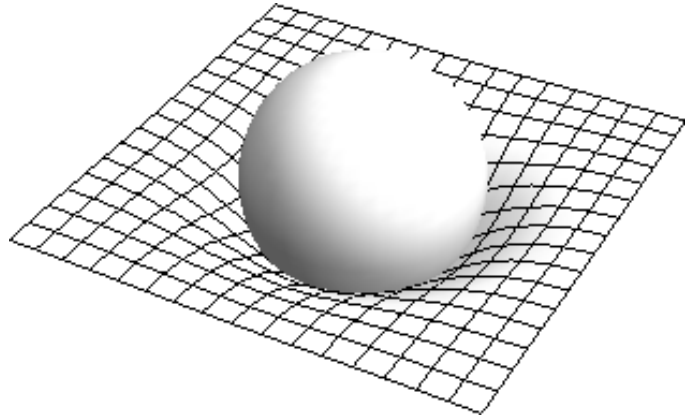}
\caption{\small{In general relativity a large mass curves the spacetime around it.}}
\label{embed}
\end{center}
\end{figure}

A fundamental underpinning of Einstein's General Theory of Relativity (GR) is the idea that the geometry of space and time is altered by the matter and energy within it. Freely falling objects follow geodesics within this curved geometry, resulting in the apparently curved trajectories that we attribute to the force of gravity. This description of gravity is often illustrated with diagrams such as Figure \ref{embed}, which represents space as a rubber sheet that is distorted by the presence of large masses. Figure \ref{embed} is an example of an embedding diagram, which shows the properties of a curved space by illustrating it as a curved surface within a higher dimensional Euclidean space.

\begin{figure} [h]
\begin{center}
\includegraphics[width = 90mm]{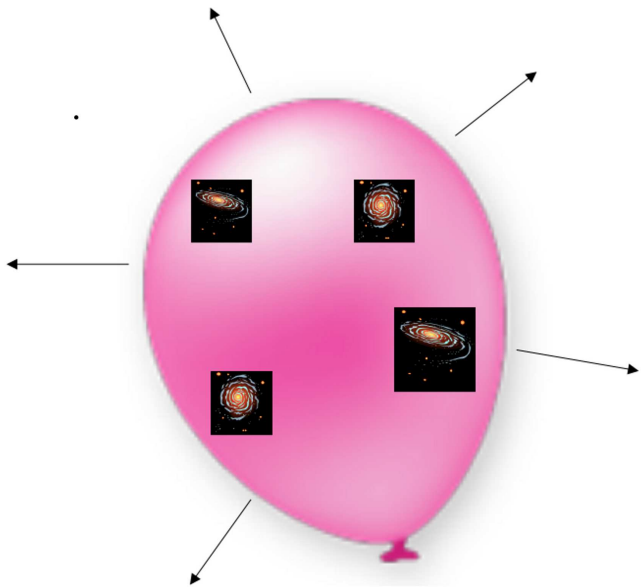}
\caption{\small{The balloon model of an expanding universe}}
\label{balloon}
\end{center}
\end{figure}

While such diagrams can be helpful, they risk giving the impression that curved geometry is an illusion and that the fundamental geometry of space must be Euclidean. There are several limitations to this viewpoint. Firstly, not all curved spatial geometries can be represented through embedding diagrams in this way. Also, this method is useless for understanding curved spacetimes. Finally, even in cases where the spatial geometry can be represented through embedding, it can lead to serious conceptual errors. For example, a homogeneous, expanding universe with constant positive curvature is often represented as the surface of an expanding balloon, as shown in Figure \ref{balloon}. Such images inevitably lead to questions about what is interior to and exterior to the balloon, and they reinforce the mistaken notion that the Big Bang took place at a particular spot: i.e. the “center of the universe.”

The goal of the CurvedLand applet is to illustrate the meaning of curved geometry without the use of embedding diagrams. The applet shows a two- dimensional world of constant positive curvature. The user can populate this world with objects of different shapes at different locations. Viewed from any one perspective, the geometry of the world is not visible, but it becomes apparent when the viewer moves through the space and observes how trajectories and shapes appear distorted. The CurvedLand applet, along with complete documentation and tutorials, is available at:
\newline
http://www.smith.edu/physics/felder/curvedland/index.html

A similar program was implemented by Coomber, et al for negative \cite{hyperbolic} and positive \cite{elliptic} curvature spaces using the LOGO Turtle Graphics environment. Their implementations focused on the curved path of the user through the space, as seen by a trail left behind the user. Our applet allows such a trail to be created but adds the ability to populate the space with points and shapes. The apparent curved motions of these objects and distortions of these shapes illustrate some of the most important aspects of non-Euclidean geometry. Our applet also provides a Web-based interface that allows the user to navigate by simply pressing control buttons rather than entering commands. In recognition of these earlier programs we made the user icon in CurvedLand a turtle.

The rest of the paper describes the features and interface of the CurvedLand applet and explains the calculations performed by the applet. The conclusions describe possible directions for future expansions of the project. We note that the name “CurvedLand” is intended as an homage to Edwin Abbot's novel Flatland \cite{flatland}, which discusses embedding of spaces within higher dimensional spaces from the point of a view of a two-dimensional character discovering the third dimension.

\section{The CurvedLand Applet}

\begin{figure} [h]
\begin{center}
\includegraphics[width = 140mm]{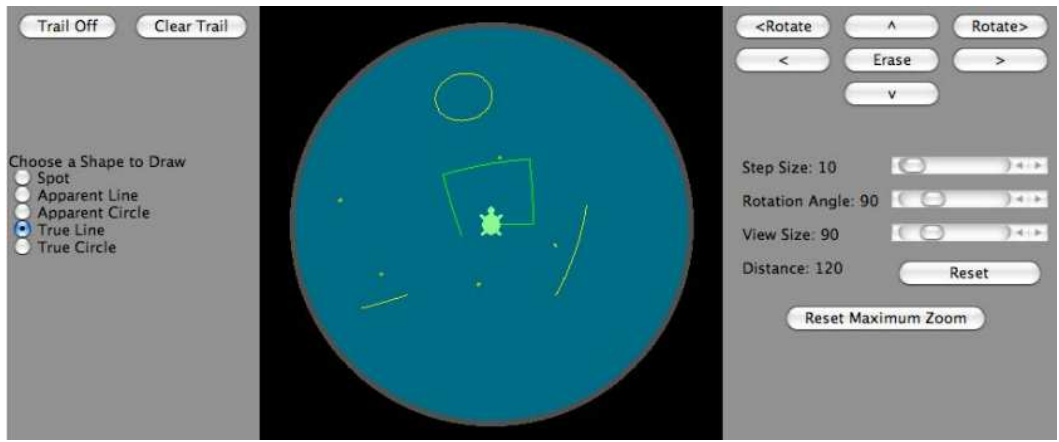}
\caption{\small{A screen shot from the CurvedLand applet, showing a combination of dots, shapes, and the trail left by the turtle.}}
\label{screenshot}
\end{center}
\end{figure}

Figure \ref{screenshot} shows a screenshot from the CurvedLand applet. The location of the user is represented by the turtle icon, which is always at the center of the screen. All other objects in CurvedLand are shown as they would be seen by the turtle. The viewer can use the mouse and radio buttons to place different shapes anywhere in the space. The user can move in CurvedLand with arrow keys that function like the arrow keys in online mapping software such as MapQuest. Picture looking at a map online and pressing a button to move North. What you see is that every object on the map appears to move South. If CurvedLand were a Euclidean space then it would behave in the same way. Instead, because of the curvature of the space, objects appear to move along curved paths as the user walks in a straight path through CurvedLand. Because different points follow different apparent paths the shapes of objects will also appear to change as the viewer moves.

\begin{figure} [h]
\begin{center}
\includegraphics[width = 30mm]{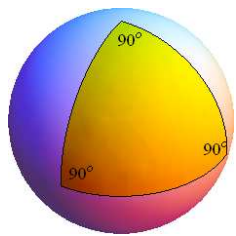}
\caption{\small{A triangle with three 90° angles in a space of positive curvature.}}
\label{triangle}
\end{center}
\end{figure}

The space represented in CurvedLand is a two-dimensional space of constant positive curvature, which is mathematically equivalent to the surface of a sphere. Although the goal of the applet is to show the effects of curvature without the use of embedding, it can nonetheless be useful to picture the surface of a sphere in order to understand the behavior of objects within the space. For example, if you zoom in so that you are looking at only a small fraction of the sphere then everything appears to behave as it would in a flat space.  This is why people can walk around on the surface of the Earth and not notice its curvature. If you zoom far out, however, the curvature quickly becomes apparent. Suppose, for example, that you consider your position to be at the North Pole and you zoom out enough to see an object on the equator directly to your right. If you walk forwards this object will remain directly to your right at a constant distance from you; from your perspective it will not appear to move at all. If you walk forward far enough to reach the equator, turn right ninety degrees, walk until you reach this object, and turn right ninety degrees again, you will find yourself facing your original location. If you walk back to the North Pole you will have traced out a triangle with three ninety-degree angles, clearly showing the non-Euclidean geometry of the space. This triangle is illustrated in Figure \ref{triangle}.

In addition to showing the apparent positions of objects in CurvedLand, the applet provides another way for you to see the geometry. The user can turn on a trail that the turtle leaves behind. With the trail on, for example, you can trace out the triangle described in the previous paragraph and see all of the legs. As you walk along any one of them, that leg will appear straight to you, but looking back at the other legs they will appear to be curved.

There are many other experiments you can try, which illustrate the properties of curved geometry. The tutorials on the CurvedLand website describe some of these, but the user can also create many more.

\section{The Calculations}

The equations used by CurvedLand to move objects and distort shapes and trajectories are most easily derived by treating CurvedLand as the surface of a sphere. Your current location is always taken to be the North Pole. The locations of points within the world are stored as longitude and latitude values. This can easily be translated into coordinates on the screen since the longitude is simply the azimuthal angle to the object and the latitude is directly proportional to its distance from the center of the screen (the turtle icon). The longitude and latitude, along with the radius of the sphere, form a triplet of spherical coordinates for the location of the object in the three-dimensional embedding space.

When the user takes a step the program converts these spherical
coordinates into cylindrical coordinates, where the principal axis of
the cylindrical coordinate system is the axis of rotation for that
step. For example, if the turtle takes a step towards the x-axis (from
its position on the z-axis) that is equivalent to rotating the sphere
around the y-axis, leaving the turtle at the North Pole. In these
cylindrical coordinates the rotation is merely a change in the angular
variable. After these changes are calculated the coordinates are
converted back to longitude and latitude.

All objects such as circles, lines, and the turtle's trail are stored as collections of points; the algorithm described above is applied separately to each point. When these objects are displayed the points are connected with lines. 

The applet allows two types of shapes to be drawn: true and apparent. True shapes are defined by the common geometrical definitions; in other words, true lines are geodesics, and true circles consist of all of the points at some distance from some center point. In contrast, apparent shapes are shapes that appear to be the Euclidean equivalents of true shapes from the viewer's current position. In general, true and apparent shapes are not the same. 

Apparent shapes are created by first finding the Cartesian coordinates
of a collection of points that represent that shape as it is displayed
on the screen. These Cartesian screen coordinates are then converted
to latitude and longitude values on the surface of the sphere using
the current view size. The shape is stored as an array of point
objects; each point consists of a latitude and longitude. True shapes
are constructed in a similar way. Because a true circle is equivalent
to an apparent circle when the viewer is at the center of the circle,
and a true line is equivalent to an apparent line at any point along
that line, the viewer is simply moved to one of those points, an
apparent shape is created, and then the viewer is moved back to its
original position. These movements are not displayed.

\section{Conclusions}

One of the first concepts that a student must grapple with in studying Einstein's general theory of relativity is the notion of curved geometry. This idea is most commonly illustrated using embedding diagrams. As we have noted above, such diagrams can reinforce or even create conceptual misunderstandings, most notably the idea that curvature is really a result of being on a curved surface, and that the underlying geometry is actually Euclidean. Many students fail to understand that a curved space can exist without a higher dimensional embedding space. CurvedLand attempts to address this misconception by showing the properties of a curved space as seen by the inhabitants of that space. We assume that most students have experience with mapping software and can thus easily appreciate how the objects in CurvedLand appear to move differently from objects in a Euclidean map.

We view the current form of the applet as a beginning. In the future
we would like to implement an option for a space of constant negative
curvature, and ultimately allow for customized, inhomogeneous
geometries. The latter may require a form of ray tracing for geodesics
to determine sight lines between objects in the space. Eventually we
would like to use three dimensional graphics to implement a first
person view in a curved, three dimensional space. Even in its current
form, however, we expect this to be a useful tool for demonstrating
one of the most challenging and important aspects of mathematical
physics.

\section{Acknowledgements}

We would like to thank the students who helped contribute to the writing of the program: Lauren Barth-Cohen, Dooshaye Moonshiram, and Nicole Brynes. We would also like to thank Gordon McNaughton for his help with Java and Web related issues. We would like to thank John Erickson for useful feedback on the website. This work was supported by NSF grant \#$0757746$.

\bibliography{curvedlandbib}

\end{document}